\documentclass[Physsubmission, Phys]{SciPost}
\hypersetup{
    colorlinks,
    linkcolor={red!50!black},
    citecolor={blue!50!black},
    urlcolor={blue!80!black}
}

\usepackage[bitstream-charter]{mathdesign}
\usepackage{caption}
\usepackage{subcaption}
\usepackage{wrapfig}

\DeclareSymbolFont{usualmathcal}{OMS}{cmsy}{m}{n}
\DeclareSymbolFontAlphabet{\mathcal}{usualmathcal}

\begin{document}

\begin{center}{\Large \textbf{
Quantized fragmentation of three-dimensional QCD string\\
}}\end{center}

% TODO: write the author list here. Use initials + surname format.
% Separate subsequent authors by a comma, omit comma at the end of the list.
% Mark the corresponding author with a superscript *.
\begin{center}
\v{S}\'{a}rka Todorova-Nov\'{a}\textsuperscript{$\star$}
\end{center}

\begin{center}
{\bf} IPNP Charles University, Prague, Czech Republic
\\
* sarka.todorova@cern.ch
\end{center}

\begin{center}
\today
\end{center}

% For convenience during refereeing (optional),
% you can turn on line numbers by uncommenting the next line:
%\linenumbers

\definecolor{palegray}{gray}{0.95}
\begin{center}
\colorbox{palegray}{
  \begin{tabular}{rr}
  \begin{minipage}{0.1\textwidth}
    \includegraphics[width=30mm]{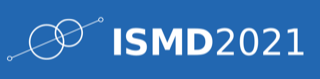}
  \end{minipage}
  &
  \begin{minipage}{0.75\textwidth}
    \begin{center}
    {\it 50th International Symposium on Multiparticle Dynamics}\\ {\it (ISMD2021)}\\
    {\it 12-16 July 2021} \\
    \doi{10.21468/SciPostPhysProc.?}\\
    \end{center}
  \end{minipage}
\end{tabular}
}
\end{center}

\section*{Abstract}
{\bf
% TODO: write your abstract here.
Recent developments of the model of quantized helical QCD string
are presented, notably the baryon production. An overview
of the experimental evidence is discussed as well as possible applications.
}

\vspace{10pt}
\noindent\rule{\textwidth}{1pt}
\tableofcontents\thispagestyle{fancy}
\noindent\rule{\textwidth}{1pt}
\vspace{10pt}

\section{Introduction}
\label{sec:intro}
 The work presented here is the latest development on a subject
 presented regularly in this conference. The very first time we heard of helical QCD string, the
 conferences venue was the ancient site of Delphi, Greece~\cite{ismd_delphi}, and
 Alessandro De Angelis was presenting DELPHI results~\cite{delphi_screwiness} disproving
 the idea by measuring the observable suggested by authors of~\cite{lund_helix}, which was freshly out of print at that time.
 I remember vividly Prof. Bo Andersson from Lund, present in the
 audience, complaining about the right of a theorist to enjoy at least
 a half an hour of glory before being disproved. The untimely death
 of Prof. Andersson in 2002 and the gap between LEP and LHC datataking
 delayed further work but by 2012, we had the experimental
 confirmation of azimuthal ordering of hadrons which could be attributed
 to underlying helical field~\cite{ATLAS-AO}, using slightly modified
 observables and minimun bias data from LHC. The model got a
 significant boost after it was noticed that three-dimensional string
 allows to explore causal connections between string break-up
 vertices, opening a wide range of model building opportunities\cite{qhelix}.

\section{Induced gluon splitting} 
    
\begin{figure}[h]
\centering
\includegraphics[width=0.4\textwidth]{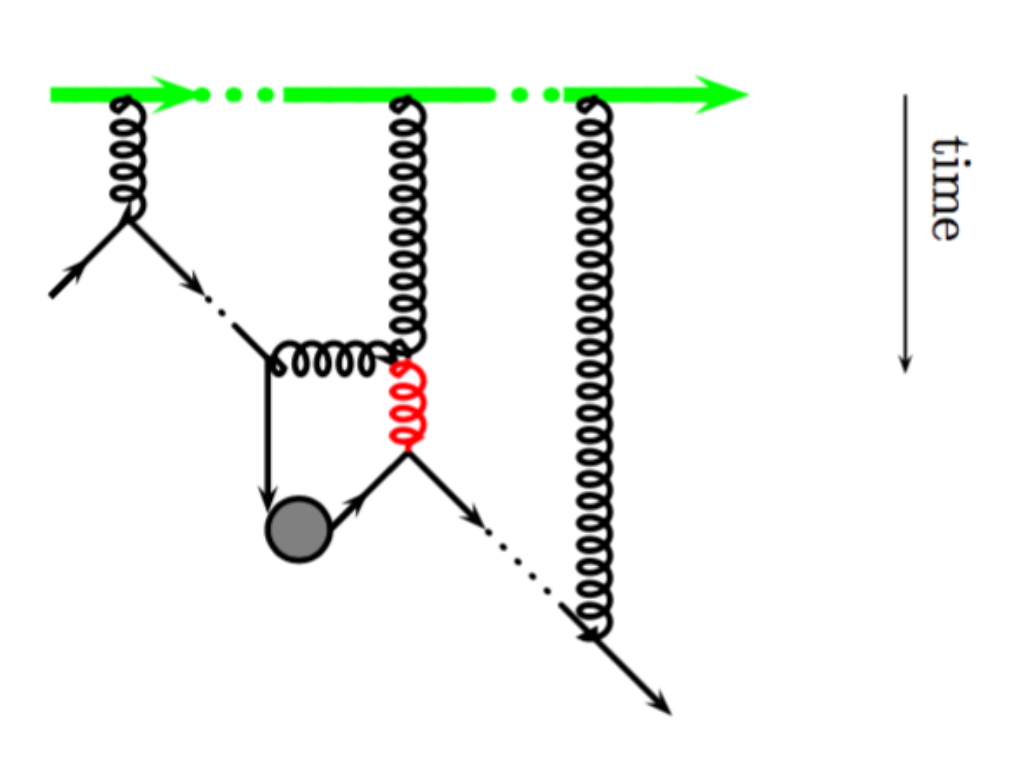}
\includegraphics[width=0.5\textwidth]{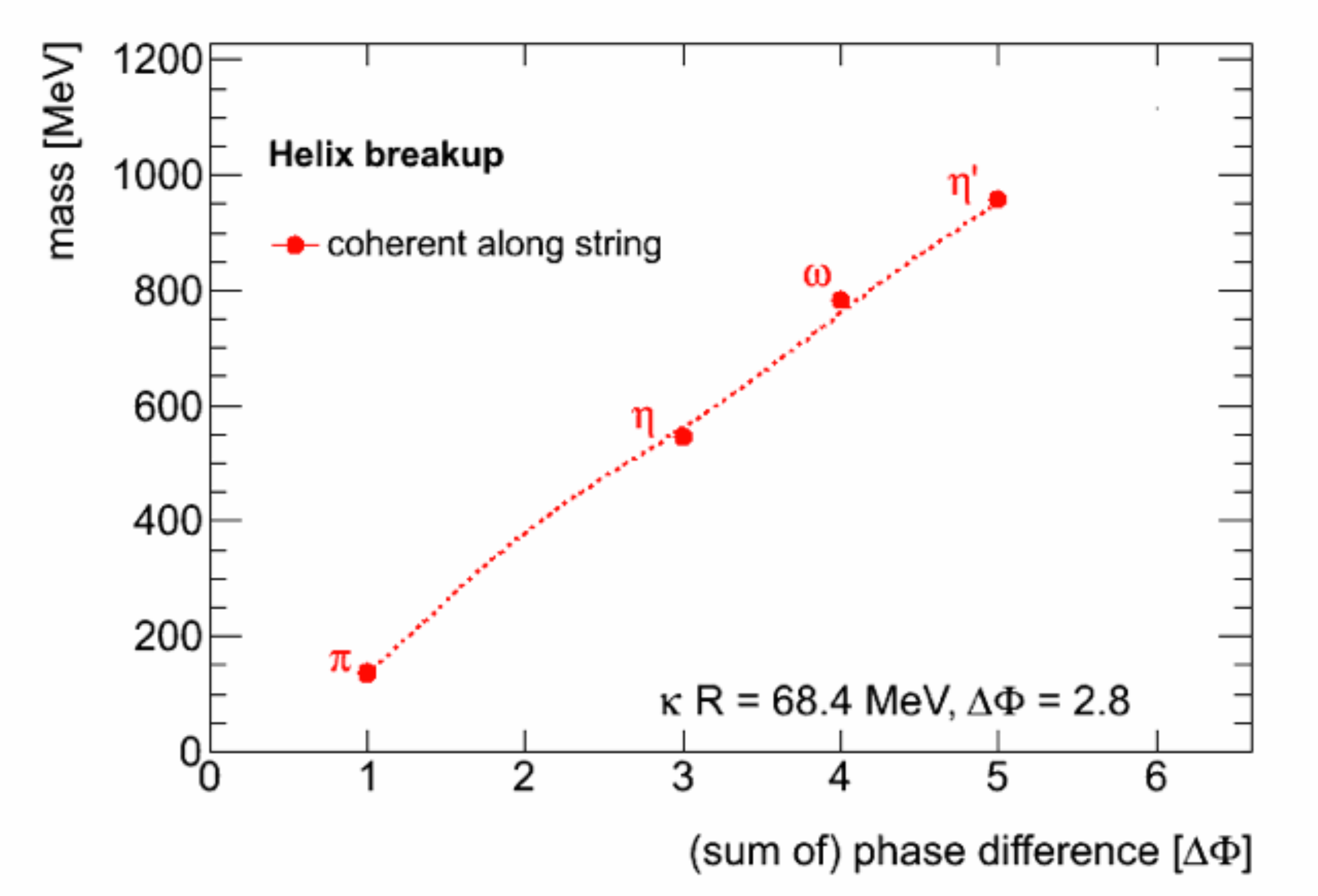}
\caption{ Left: Scheme of causal (induced) gluon splitting - the
  excited gluon marked in red decays promptly into a $Q\bar{Q}$
  pair. Right: Causal approach to string breaking reveals the
  quantized nature of the process. Plots taken from \cite{baryons}.
\label{fig:light-front}}
\end{figure}

  Let's consider a case of a massless quark - or antiquark - pulled by the string
  tension tangential to the (helical) string in close analogy with
  standard ( one-dimensional ) Lund string~\cite{lund}. As the quark propagates
  along the string with the speed of light, it acquires a longitudinal
  and transverse momentum component, and the curvature of the helical
 string translates part of the accumulated transverse energy into
 (dynamic) quark
 mass.  In this case, there is just one possible scenario for causally
 connected string vertices, and that's the one in which the quark
 triggers a split of gluon ($g\rightarrow q\bar{q}$). This
 ``light-front'' string fragmentation has another interesting property
 and that is the decoupling of the transverse and longitudinal momenta
 components, where the mass and the intrinsic transverse momentum of
 the resulting hadron is entirely driven by the transverse shape of
 the string ( a constant string tension $\kappa$=1GeV/fm is used in
 the model, as in the standard Lund fragmentation ).
Such a hadron
 production scenario is suitable for creation of hadrons with limited
 internal degrees of freedom (lightest hadrons).  The lightest hadrons
 decaying into (1,3,5) pions are pseudoscalar mesons
 ($\pi,\eta,\eta'$) and comparison of their mass spectra with the
 fragmentation of helical string reveals a quantized fragmentation
 pattern where string splits in phase intervals $\Delta\Phi\sim$2.8
 rad, with $\kappa R\sim$0.07GeV, $R$ standing for the radius of
 helix (Fig.\ref{fig:light-front}).  The scheme also fits formation of $\omega$ meson, a vector
 meson which takes 4 quanta of helical string and decays into n$<$4
 pions (3,preferably). 

Although the model is very simple and does not take into account mass
of quarks, it nevertheless describes the mass spectrum with
the precision of up to 3\%, and inversely, the mass spectrum
constraints the string parameters $\Delta\Phi, \kappa R$ with similar precision.

\section{Baryon production}

   For a closely packed helix string winding (small pitch), one can
   imagine the induced gluon splitting can take place across string
   loops as shown in Fig.~ref{fig:across-loops}.  A sequence of two
   induced breakups of this type produces coherent systems of 3 quark
   and 3 antiquarks which may emerge as baryonic states. It turns out
   the lightest baryons (nucleons) fit within the scheme of quantized
  helical string fragmentation quite naturally, as n=5 states. There is no need to
  introduce new parameters in the model, nor adjust the string
  parameters, and nucleon mass is reproduced with 1\% precision \cite{baryons}.
  
\begin{figure}[h]
\centering
\includegraphics[width=0.5\textwidth]{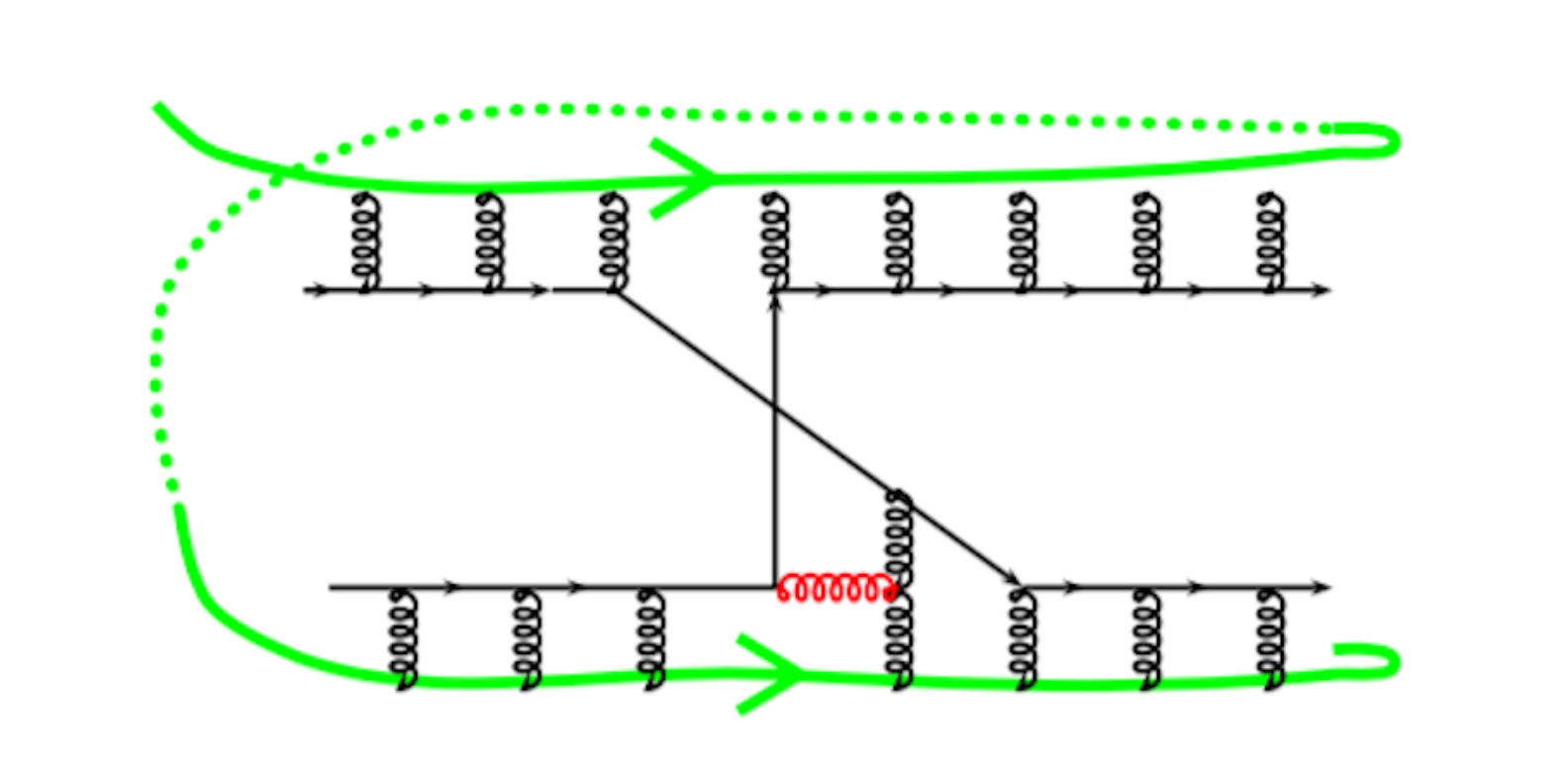}
\includegraphics[width=0.45\textwidth]{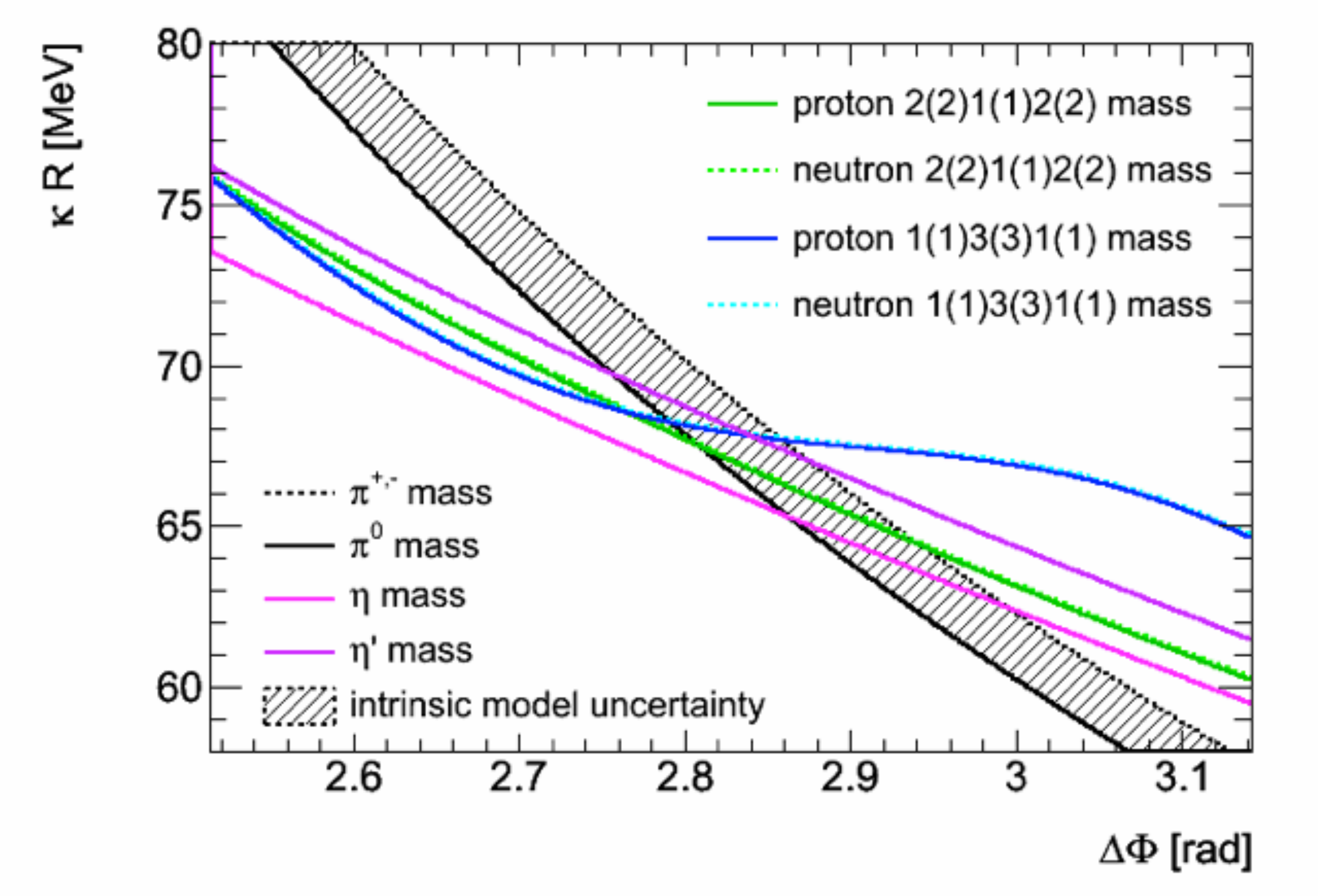}
\caption{ Left: Scheme of causal (induced) gluon splitting across
  string loop - the
  excited gluon marked in red decays promptly into a $Q\bar{Q}$
  pair. Right: Relation between string parameters and hadron mass. Nucleon mass corresponds to expectation of the
  quantized fragmentation scheme, within model uncertainties. Plots taken from \cite{baryons}.
\label{fig:across-loops}}
\end{figure}
 
%\begin{wrapfigure}{r}{0.5\textwidth}
\begin{figure}[h]
\centering
\includegraphics[width=0.6\textwidth]{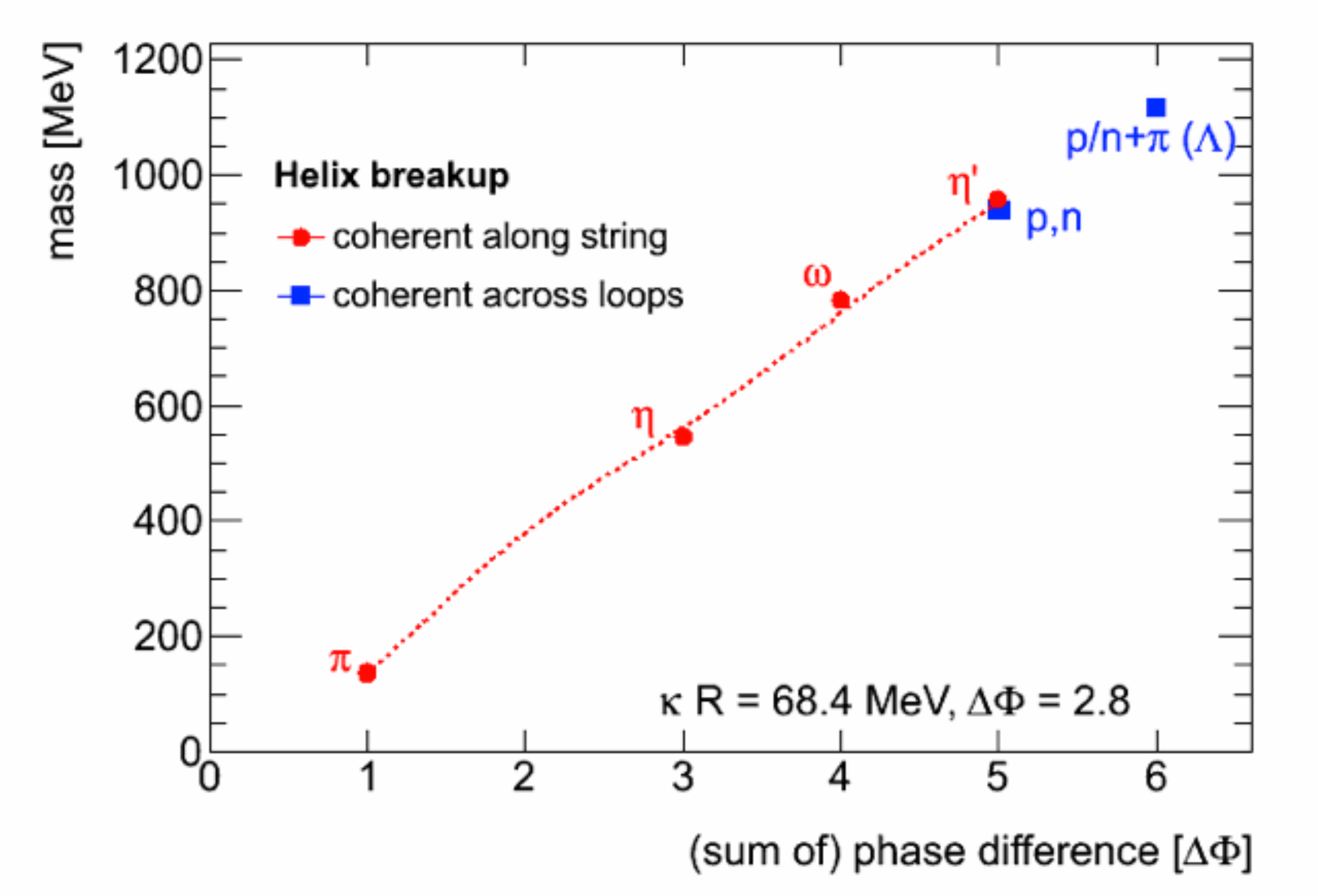}
\caption{ Spectrum of light mesons and baryons matching the scheme of
  quantized string fragmentation via induced (causal) string breakup. Plot taken from \cite{baryons}.
\label{fig:lambda}} 
\end{figure}
%\end{wrapfigure}

 It is interesting to note \cite{baryons}  that not only $\Lambda$
 baryon mass is obtained from n=6 state modelling without further
 adjustements,
 but also as an unbound p+$\pi$ state  ends up with mass equivalent
 to $\Lambda^0$ resonance in the quantized fragmentation of helical
 string with parameters constrained by mass of pseudoscalar mesons.
  This raises a possibility that part of  $\Lambda$ baryon
   production does not follow the established quark model description
   and may not imply a strange quark creation.

\section{Correlations}
\begin{figure}[h]
\centering
\includegraphics[width=0.5\textwidth]{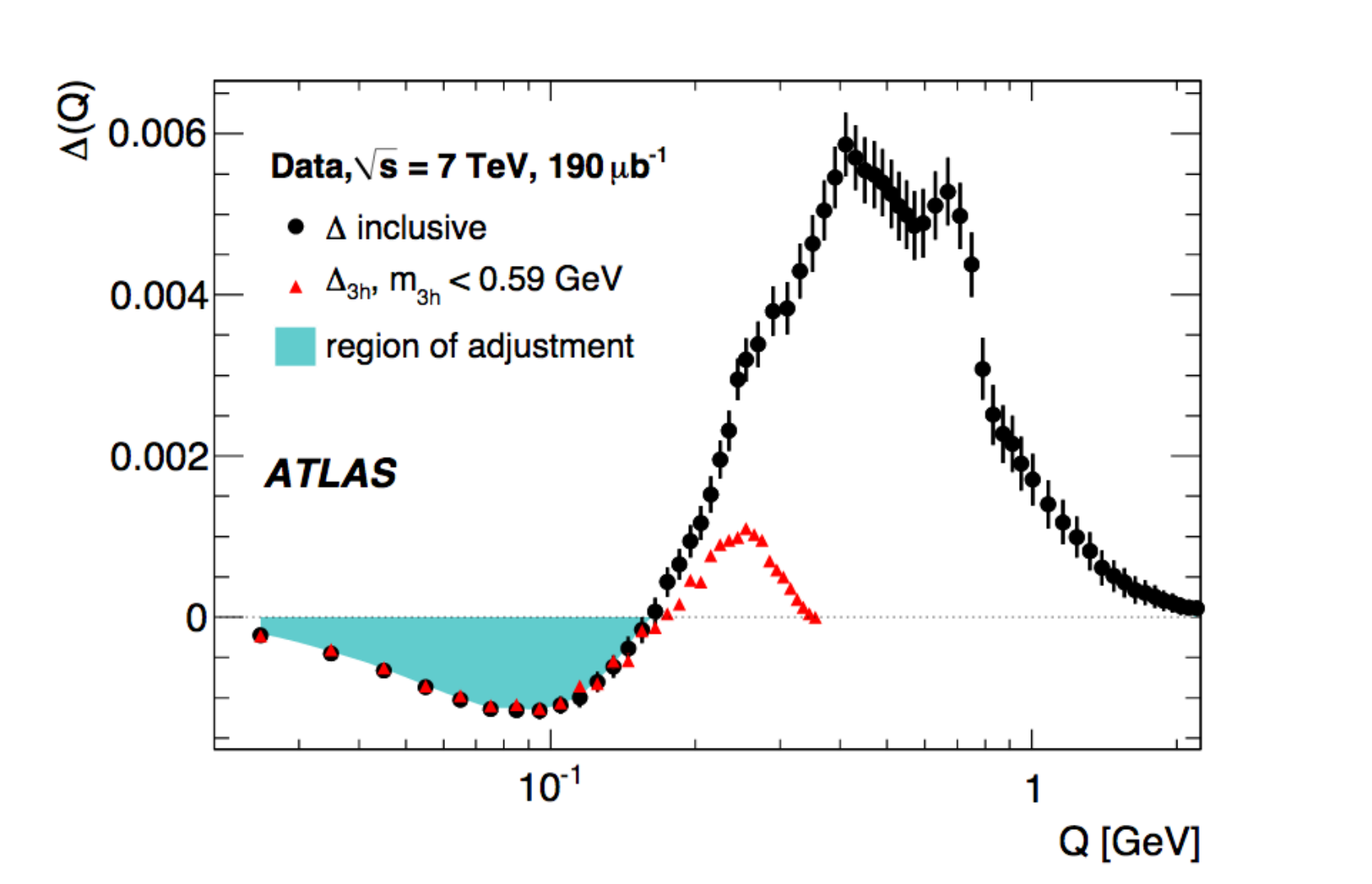}
\includegraphics[width=0.45\textwidth]{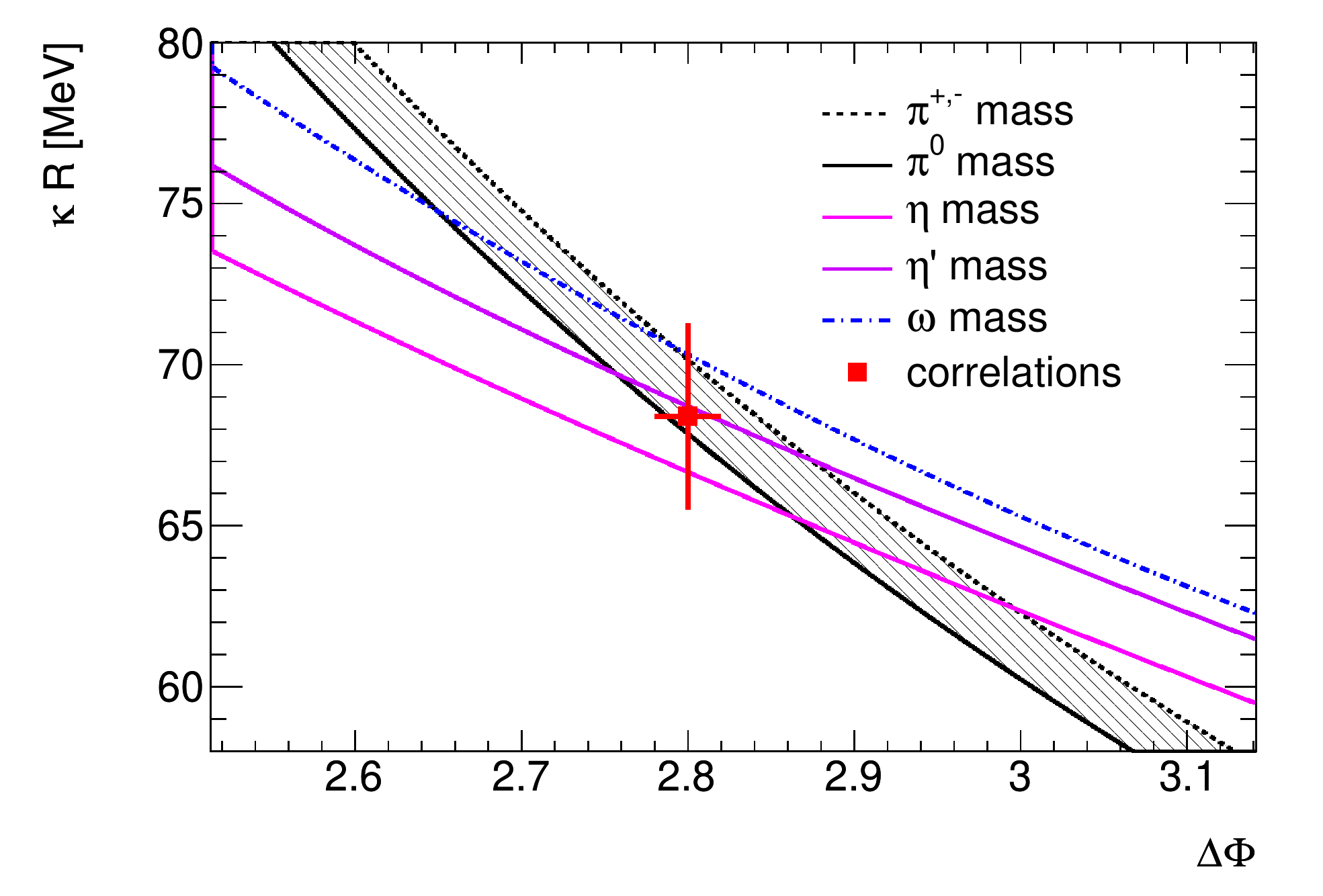}
\caption{ Left: ATLAS measurement of ordered hadron chains. The
  turquoise region marks the excess of like-sign pairs traditionally
  attributed to Bose-Einstein interference~\cite{ATLAS-chains}. Right:
  Measurement of correlations provides an independent cross-check of
  string parameters, in good agreement with values obtained from study
  of hadron mass spectra~\cite{baryons}.   
\label{fig:correlations}}
\end{figure}

    The fact that quantization proceeds in transverse energy
    $E_T=\sqrt{m^2 +p_T^2}$ (with respect to string axis), rather than
    mass itself,  implies that on top of hadron masses, also the
    intrinsic transverse momenta of hadrons are quantized, and
    correlations between adjacent hadrons predicted, at least for the
    case of helical string with constant pitch. For a chain of direct
    pions, the model predicts higher momentum difference for a pair of
    pions with rank difference 1 (adjacent) than for a pair with rank
    difference 2 ( rank describes ordering of hadrons according to
    colour flow). In combination with the local charge conservation
    which forbids a creation of like-sign adjacent pions, the model
    predicts charge-asymmetry in rank=1,2 production, a threshold-like
    momentum difference for adjacent (unlike-sign) pion pairs
    (Q$\sim$0.26 GeV) and 
    an excess of like-sign pairs at low Q$\sim$0.09 GeV). The
    measurement has been performed by ATLAS~\cite{ATLAS-chains}. The
    signature of ordered hadron chains with predicted properties was found
    and it was established that the measured chains carry the entire
    like-sign pair enhancement traditionally attributed to
    Bose-Einstein interference.  The measurement of the shape of
    correlations within ordered chains provides an independent
    evaluation of string parameters~\cite{baryons}, and a good
    agreement is observed between results derived from study of hadron
    spectra and from correlations (Fig.~\ref{fig:correlations}).

\section{Conclusion}
   The most natural explanation for the quantization of QCD field is
   that there is a limited number of gluons in the field, perhaps as
   few as two gluons per $\Delta\Phi$. This would mean the
   quantization, on which we may have quite a good experimental
   handle, wipes away the sea of non-perturbative gluons. Instead, we
   have a well ordered, sparsely populated QCD vacuum, which
   brings us back to the subject of the original ``screwiness'' paper~\cite{lund_helix}.  

 \begin{figure}[h]
\centering
\includegraphics[width=0.6\textwidth]{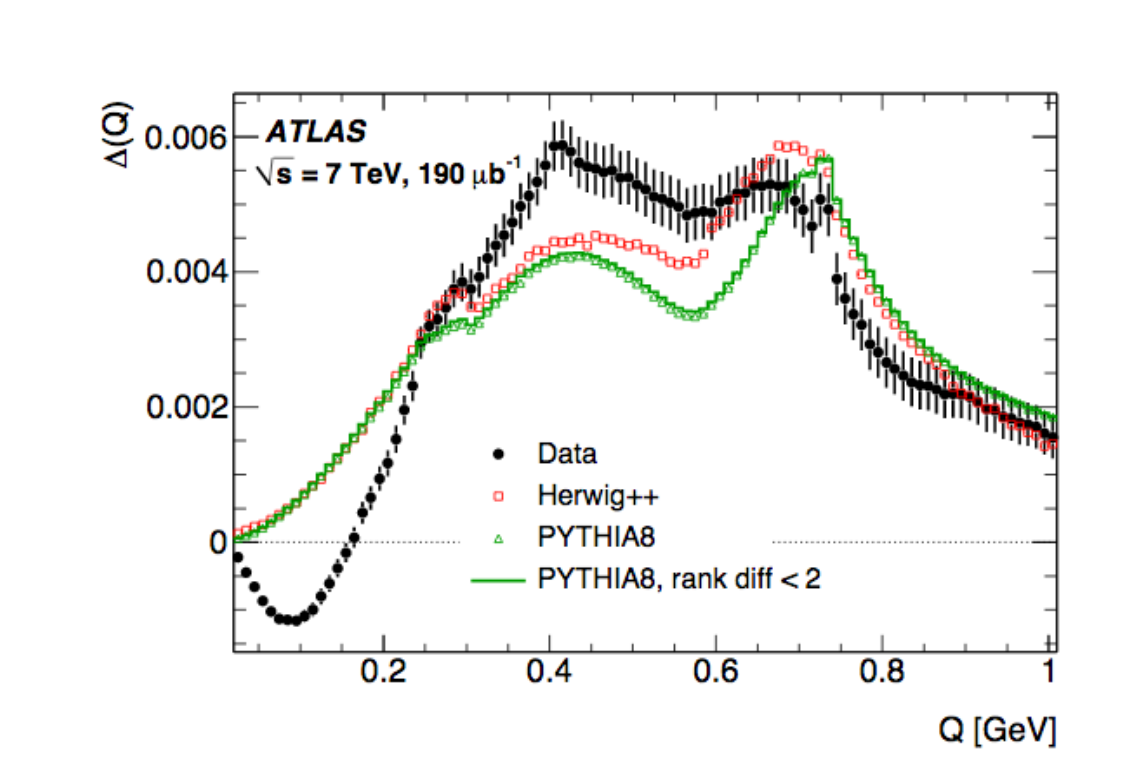}
\caption{Conventional hadronization models fail to describe data, and
  the lack of quantum effects seems to be the primary source of discrepancies. Plot taken from \cite{ATLAS-chains}.
\label{fig:dq}} 
\end{figure}

    Currently, the particle physics community does not have a reliable
    tool for the study of hadronization systematics, and the lack of
    quantum effects may well be the main source of discrepancies.
    Fig.~\ref{fig:dq} illustrates the situation with the ATLAS measurement of
   the  difference between inclusive distribution of unlike-sign and like-sign hadron pairs 
   ( normalized to the number of charged particles in the sample),
   which reflects the resonance decay and correlations between
   adjacent hadrons~\cite{ATLAS-chains}. Both conventional Lund fragmentation and the
   clusterization fail to describe the data in a similar way, because
   none of them takes into account natural quantum thresholds and other quantum effects.

\section*{Acknowledgements}
This work is supported by the Inter-Excellence/Inter-Transfer grant LT17018 and the Research Infrastructure project LM2018104
 funded by Ministry of Education, Youth and Sports of the Czech Republic, and the Charles University project UNCE/SCI/013.

% TODO:
% Provide your bibliography here. You have two options:

% FIRST OPTION - write your entries here directly, following the example below, including Author(s), Title, Journal Ref. with year in parentheses at the end, followed by the DOI number.
%\begin{thebibliography}{99}
%\bibitem{1931_Bethe_ZP_71} H. A. Bethe, {\it Zur Theorie der Metalle. i. Eigenwerte und Eigenfunktionen der linearen Atomkette}, Zeit. f{\"u}r Phys. {\bf 71}, 205 (1931), \doi{10.1007\%2FBF01341708}.
%\bibitem{arXiv:1108.2700} P. Ginsparg, {\it It was twenty years ago today... }, \url{http://arxiv.org/abs/1108.2700}.
%\end{thebibliography}

% \bibliographystyle{SciPost_bibstyle} % Include this style file here only if you are not using our template
%\bibliography{SciPost_Example_BiBTeX_File.bib}

\nolinenumbers

\end{document}